 \documentclass[pdflatex,sn-nature]{sn-jnl}

\usepackage{graphicx}%
\usepackage{multirow}%
\usepackage{amsmath,amssymb,amsfonts}%
\usepackage{amsthm}%
\usepackage{mathrsfs}%
\usepackage[title]{appendix}%
\usepackage{xcolor}%
\usepackage{textcomp}%
\usepackage{manyfoot}%
\usepackage{booktabs}%
\usepackage{algorithm}%
\usepackage{algorithmicx}%
\usepackage{algpseudocode}%
\usepackage{listings}%
\usepackage{float}%
\usepackage{enumitem}
\usepackage{rotating}
\usepackage{subfigure}

\theoremstyle{thmstyleone}%
\newtheorem{theorem}{Theorem}
\theoremstyle{thmstyletwo}%
\algnewcommand{\PreDetectionEncoding}{\item[\textbf{Pre-Detection and Encoding:}]}
\algnewcommand{\Input}[1]{\item[\textbf{Input:}] #1}  
\algnewcommand{\Output}[1]{\item[\textbf{Output:}] #1}
\algnewcommand{\Preprocessing}[1]{\item[\textbf{Preprocessing:}] #1}

\theoremstyle{thmstylethree}%
\newtheorem{definition}{Definition}%

\begin{document}

\title[Article Title]{Efficient Maximum Clique Detection via Grover’s Algorithm with Real-time Global Size Tracking}


\author[1,2]{\fnm{Wenmin} \sur{Han}}

\author[1,2]{\fnm{Shiqi} \sur{Zheng}}
\author[1,2]{\fnm{Peian} \sur{Chen}}
\author*[1,2]{\fnm{Yukun} \sur{Wang}}\email{wykun06@gmail.com}

\affil[1]{\orgdiv{Beijing Key Laboratory of Petroleum Data Mining}, \orgname{China University of Petroleum}, \orgaddress{\city{Beijing}, \postcode{102249},  \country{China}}}

\affil[2]{\orgdiv{Key Lab of Processors, Institute of Computing Technology}, \orgname{CSA}, \orgaddress{\city{Beijing}, \postcode{100190}, \country{China}}}

\abstract{The maximum clique problem(MCP) is to find the largest complete subgraph in an undirected graph, that is, the subgraph in which there are edges between every two different vertices. It is an NP-Hard problem with wide applications, including bioinformatics, social networks, data mining, and other fields. This paper proposes an improved algorithm that dynamically tracks the maximum clique size by encoding prior constraints on the vertex count—derived from Turán's theorem and complete graph properties—into global variables through quantum circuit pre-detection. The algorithm further synergizes with Grover's search to optimize the solution space.  
Our auxiliary-qubit encoding scheme dynamically tracks clique sizes during quantum search, eliminating iterative measurements, achieving MCP solution with $O\left ( \sqrt{2^{n } }  \right ) $ Grover iterations and $O\left ( 1 \right ) $ measurements. This represents an $\boldsymbol{n}$-fold improvement over state-of-the-art Grover-based methods, which require $O(n\sqrt{2^n})$ iterations and $O(n)$ measurements for $n$-vertex graphs. We validate algorithmic correctness through simulations on IBM's Qiskit platform and benchmark qubit/gate efficiency against existing Grover-based MCP solvers.}


\maketitle

\section{Introduction}\label{intro}

The maximum clique problem is to find the largest complete subgraph in a given undirected graph. Its decision version is one of the original 21 NP-complete problems established by Karp in his seminal paper on computational complexity \cite{karp2009reducibility}. As a fundamental problem in combinatorial optimization, the MCP has been extensively studied \cite{pardalos1994maximum,bomze1999maximum,eblen2012maximum}
and finds wide application across diverse domains. These include its significant role in social network analysis \cite{adraoui2022maximal}, as well as applications in bioinformatics \cite{tomita2011efficient}, data mining \cite{wang2009order}, and communication signal processing \cite{douik2020tutorial}, among others. 

Due to its fundamental theoretical significance and broad practical utility, MCP has attracted considerable attention from both academia and industry, spurring the development of numerous solution algorithms. Currently, the classic algorithms for solving MCP primarily fall into three categories:  exact algorithms based on the Branch-and-Bound {(B\&B) method \cite{carraghan1990exact}}
, {heuristic algorithms that adopt  approximation strategies \cite{wang2016two}},%
and  deep learning-based methods utilizing Graph Neural Networks\cite{wu2015review} \cite{karalias2020erdos}\cite{bomze1999maximum}\cite{marino2024short}.
The most mainstream exact algorithms are based on the branch-and-bound method, which estimates the upper bound through a graph coloring procedure, with a worst-case time complexity of $O\left ( 3^{\frac{n}{3} }  \right ) $
\cite{wu2015review,tomita2017efficient,marino2024short},which grows rapidly with the number of vertices 
$n$ in the graph. 
As a result, exact solutions become computationally infeasible for large-scale graphs. Furthermore, obtaining good approximation ratios in polynomial time is also extremely challenging. 
In fact,  unless NP = ZPP, for any $\varepsilon > 0$, the Maximum Clique Problem cannot be approximated within a factor of $n^{1-\varepsilon } $ in polynomial time\cite{johan1999astad}.

\begin{figure}[h]
    \centering
    \includegraphics[width=0.9\textwidth]{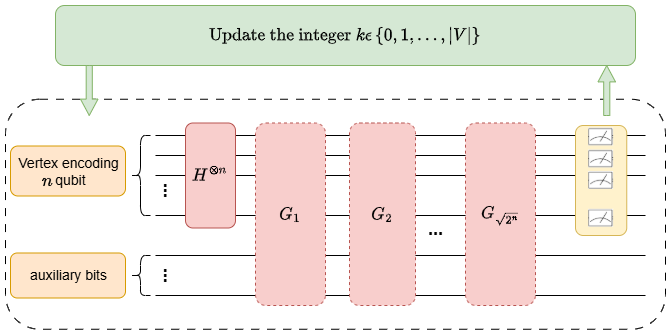}
    \caption{In the existing framework for solving the maximum clique problem using the Grover algorithm, the parameter $k$ is used to mark and amplify the cliques whose number of vertices is greater than $k \epsilon \left [ 1,\left | V \right |  \right ]$ in each iteration.The value of parameter $k$ is dynamically adjusted according to the measurement results in each iteration, gradually approaching and identifying the maximum clique.
}
    \label{fig1}
\end{figure}

{Quantum computing demonstrates potential speedups for specific hard problems over classical approaches \cite{nielsen2010quantum,albash2018adiabatic,chen2021review}}
, motivating numerous quantum methods targeting the MCP. 
Childs et al.\cite{childs2000finding} proposed an adiabatic algorithm for MCP. While promising in principle, it indicates that rigorous characterization of its asymptotic performance appears challenging. Pelofske and Maji demonstrated effective large clique discovery using a D-Wave quantum annealer \cite{pelofske2019solving}, though primarily on sparse graphs. Ha et al. improved QAOA for MCP via a Difference of Convex Algorithm (DCA) warm-start \cite{ha2024solving}, outperforming both standard QAOA and GNN warm-starts \cite{jain2022graph}, but the optimal solution probability remains relatively low.Significantly, Lingxiao Li's active incremental learning framework\cite{li2025efficient} presents a promising direction for enhancing QAOA-based solutions to graph problems.
Alternative approaches include oracle-based search algorithms like Grover's algorithm \cite{chang2018quantum}, which offers a provable quadratic speedup for various NP-complete problems (e.g., minimum dominating set \cite{zhang2024quantum}, graph coloring \cite{mukherjee2022grover,wang2011improved,saha2015synthesis}, Hamiltonian cycle \cite{vidya2006applying,jiang2022quantum,jiang2023solving}) and holds promise for MCP.
However, existing Grover-based MCP implementations face critical limitations: requiring up to $O(n\sqrt{2^n})$  iterations and $O(n)$ measurements \cite{diasagrover,bojic2012quantum,chang2018quantum,sanyal2020circuit,haverly2021implementation,dias2022algoritmos}. As illustrated in  Fig.\ref{fig1}, this inefficiency stems from the quantum circuit's inability to dynamically share global clique size information during execution. Unlike classical algorithms that adaptively use vertex count $k$ to guide searches, quantum states cannot reveal intermediate clique sizes without measurement. This prevents real-time solution quality monitoring, forcing circuits to rely on a static parameter $k$ to identify cliques of size $\leq k$, which can only be updated after measurement. Consequently, each iteration requires a full Grover search with $O(\sqrt{2^n})$ operations and $O(1)$ measurements. Incrementing $k$ from minimum values toward $k_{\max} \in O(n)$ results in $\boldsymbol{O(n)}$ total measurements.

To solve this problem, we propose a Pre-Detection and Encoding strategy based on Turán's theorem and the properties of complete graphs, encodes the maximum clique size into auxiliary qubits as a dynamic global variable, replacing static $k$. Meanwhile, in the Oracle circuit, operations such as vertex copying, sorting, and XNOR comparison are employed to collaboratively filter and mark the maximum clique with the global variable for precise identification. 
The entire process requires only $O\left ( \sqrt{2^{n} }  \right ) $ Grover iterations (composed of the Oracle and diffusion operator) along with $O(1)$ measurements to deterministically retrieve all maximum cliques when the number of vertices in a graph is $n$, significantly improving the algorithm efficiency and resource utilization benefits.

The paper is structured as follows: Section \ref{preli} introduces quantum circuits, Grover’s algorithm, and the maximum clique problem fundamentals. Section \ref{imple} details the design of our proposed algorithmic circuit components and provides a concrete example demonstrating its effectiveness. Section \ref{res} rigorously analyzes quantum resource requirements (qubit count and gate complexity) and compares them with existing Grover-based maximum clique algorithms. Finally, Section \ref{con} summarizes key contributions and outlines future research directions.

\section{Preliminaries}\label{preli}
In Section \ref{preli}, we introduce the fundamental knowledge of quantum computing from the perspectives of qubits and quantum gates, and provide a brief overview of Grover's algorithm and the maximum clique problem to facilitate readers' understanding of the proposed algorithm in this paper.
\subsection{Fundamentals of Quantum Computing}
Quantum computing fundamentally involves the precise control of qubits and quantum gates. By manipulating qubit states and applying quantum logic operations, it enables information processing beyond classical computational limits.
\subsubsection{Qubits} 
Qubits, the quantum counterparts of classical bits, underpin quantum computing with their ability to exist in superposition states \(\vert\psi\rangle=\alpha\vert0\rangle+\beta\vert1\rangle\) (\(\vert\alpha\vert^2+\vert\beta\vert^2=1\)), enabling parallel computation of \(2^n\) states for n qubits. This contrasts sharply with classical bits, which are limited to binary states. Leveraging properties like superposition and entanglement, quantum algorithms achieve exponential speedup over classical counterparts in specific problems
In the quantum circuit architecture proposed in this work, qubits are functionally categorized into two distinct types based on their operational roles. The first type comprises data qubits, which serve as the primary information carriers to encode the vertex and edge data of the input graph structure. The second type consists of ancilla qubits that support computational processes by performing three critical functions: temporarily storing intermediate computational results, implementing Oracle operations, and facilitating quantum state transformations during algorithm execution. This functional classification enables efficient resource allocation and optimized circuit design for the quantum algorithm implementation.

\subsubsection{Quantum Gate} 
{Quantum gates, the fundamental logical units of quantum computing, function similarly to classical logic gates but operate on qubit states. The key quantum gates employed in this algorithm include: the Hadamard (H) gate for creating superposition states, the CNOT gate for establishing entanglement, the Toffoli (CCNOT) gate for three-qubit controlled operations, the multi-controlled Toffoli (MCT) gate for generalized control, and the Fredkin (controlled-SWAP) gate for conditional qubit exchange.}
\begin{itemize}
\item \textbf{H gate} The Hadamard gate (H gate) is a fundamental single-qubit operation that creates quantum superposition states. Its action on computational basis states can be mathematically expressed as:
\begin{displaymath} 
H\left | 0 \right \rangle=\frac{\left | 0  \right \rangle+\left | 1 \right \rangle  }{\sqrt{2} },
H\left | 1 \right \rangle=\frac{\left | 0  \right \rangle-\left | 1 \right \rangle  }{\sqrt{2} }
\end{displaymath}
As a cornerstone of quantum parallelism, the H gate is extensively employed in the initialization phase of quantum algorithms to prepare uniform superposition states.
\end{itemize}
\begin{itemize}
\item \textbf{ CNOT gate } The CNOT gate (controlled-NOT gate) is a two-qubit quantum operation that flips the target qubit when the control qubit is in the $\left | 1  \right \rangle$ state. Its operation can be mathematically represented as:
\begin{displaymath} 
CNOT\left | c,t  \right \rangle=\left | c  \right \rangle\otimes \left ( X^{c}\left | t  \right \rangle   \right )    
\end{displaymath}
\end{itemize}
\begin{itemize}
\item \textbf{ Toffoli gate } The Toffoli gate (CCNOT gate) is a three-qubit quantum operation that flips the target qubit if and only if both control qubits are in the $\left | 1  \right \rangle$ state. Its unitary transformation is mathematically expressed as:
\begin{displaymath} 
Toffoli\left | a,b,t  \right \rangle =\left | a,b  \right \rangle\otimes \left ( X^{ab}\left | t  \right \rangle   \right )    
\end{displaymath}
\end{itemize}
\begin{itemize}
\item \textbf{ MCT gate } The Multi-Controlled Toffoli (MCT) gate represents an n-qubit generalization of the standard Toffoli gate, which flips the target qubit if and only if all n control qubits are in the $\left | 1  \right \rangle$ state. In quantum search algorithms, this gate is particularly essential for constructing oracle operations that implement multi-condition evaluations.
\end{itemize}
\begin{itemize}
\item \textbf{ Fredkin gate } The controlled-SWAP (Fredkin) gate is a three-qubit quantum operation that swaps the states of two target qubits conditioned on the control qubit being in the $\left | 1  \right \rangle$ state. Its unitary transformation is mathematically expressed as:
\begin{displaymath} 
  Fredkin\left | c,a,b  \right \rangle =\left | c  \right \rangle\otimes \left ( SWAP^{c} \left | a,b  \right \rangle  \right )    
\end{displaymath}
\end{itemize}
\subsection{{Grover's algorithm}}
Grover's algorithm stands as a landmark achievement in quantum computing theory. Since its proposal by Lov Grover in 1996 \cite{grover1996fast}, researchers have significantly extended and enhanced the algorithm in various directions. {Early extensions focused on specific problems: Dürr and Høyer \cite{durr1996quantum} were among the early researchers who extended it to the problem of quantum minimum finding. Quantum counting\cite{brassard1998quantum}\cite{wie2019simpler}\cite{aaronson2020quantum}, which combines Grover's algorithm with the quantum Fourier transform, constitutes another important extension of Grover's algorithm.With the evolution of quantum technologies, Grover's operator now extends beyond initial applications to cutting-edge domains: optimizing state transfer in quantum walks \cite{inui2004localization, giri2023quantum}, enhancing AES encryption security and quantum key distribution protocols in quantum information \cite{grassl2016applying, hsu2003quantum}, and enabling quantum machine learning and quantum optimization \cite{cherif2023quantum, chakrabarty2017dynamic}.
Beyond these problem-specific applications, foundational improvements to the algorithm itself were pursued: generalization of initial states from uniform superposition to arbitrary pure states \cite{biham1999grover}, revealing success probability's dependence on initial distribution statistics (mean/variance), with subsequent extension to mixed states \cite{biham2002grover}; phase optimization through rotation techniques that achieve near-deterministic outcomes \cite{long2001grover}, complementing earlier deterministic approaches \cite{brassard2000quantum, hoyer2000arbitrary}; enhanced multi-objective search via initial state optimization \cite{byrnes2018generalized} and solution space expansion \cite{mehri2018modified}; along with advanced methods for unknown solution counts including quantum counting \cite{brassard2000quantum}, adaptive iteration \cite{boyer1998tight}, and fixed-point search employing selective phase shifts \cite{grover2005fixed}.

The generic framework of Grover's algorithm is illustrated in Fig.\ref{fig2}. The core of Grover iteration consists of two fundamental components: the oracle operation and the diffusion operator. The oracle operation marks the target state by applying a phase inversion, while the diffusion operator amplifies the amplitude of the marked state through an inversion about the mean operation.The algorithmic workflow can be distilled into four pivotal stages:
\begin{figure}[h]
    \centering
    \includegraphics[width=0.9\textwidth]{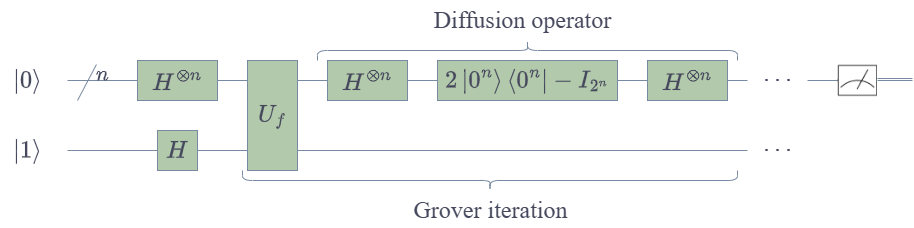}
    \caption{Generalised circuit for Grover's algorithm.}
    \label{fig2}
\end{figure}
\subsubsection{Initialization}
When searching for the marked state within the set of all computational basis states defined by $n$ qubits, the primary step of the algorithm is to prepare a $n$-qubit system in a uniform superposition of all basis states. This crucial initialization process is achieved by sequentially applying $H^{\otimes n}$ .
\begin{displaymath} 
   H^{\otimes n}\left | 0  \right \rangle^{\otimes n}   = \frac{1}{\sqrt{2^{n} } } {\textstyle \sum_{x=0 }^{2^{n}-1}}  \left | x  \right \rangle .  
\end{displaymath}
Within the algorithmic framework illustrated in Figure 2, the auxiliary qubit in the terminal position plays a critical role in the phase inversion operation of the Oracle step. This ancillary qubit is prepared in the superposition state $\left | - \right \rangle =\frac{\left | 0  \right \rangle-\left | 1  \right \rangle  }{\sqrt{2} } $ by applying a Hadamard gate to the initial $\left | 1  \right \rangle$ state. This preparation is essential for enabling the subsequent phase flip of target states during Oracle operations, thereby allowing Grover's algorithm to effectively discriminate between target and non-target states through the amplitude amplification mechanism.
\subsubsection{Oracle}
In Grover's algorithm, the role of the Oracle is to mark the target state (i.e., the solution to the search problem) by applying a negative phase to it. It is worth noting that the Oracle implements a black-box operation $U_f$, whose behavior is determined by the Boolean function $f(x)$. When $x$ is the target state, $f(x) = 1$ and 
 $U_f|x\rangle=-|x\rangle$ ; when $x$ is not the target state,$f(x) = 0$ and $U_f|x\rangle = |x\rangle$. Its simplified mathematical expression is as follows.
 \begin{displaymath} 
  U_{f}=\left ( -1 \right ) ^{f\left ( x \right ) }  \left | x \right \rangle  .  
\end{displaymath}
\subsubsection{Diffusion}
The diffusion operator amplifies the amplitude of the target state marked by the Oracle through an inversion about the mean operation, mathematically represented as $H^{\otimes n}\left ( 2 \left | 0^{ n}  \right \rangle\left\langle0^{ n}\right|-I   \right ) H^{\otimes n} $, which is equivalent to $2\left | \psi  \right \rangle \left\langle\psi\right| -I$ where $\left | \psi  \right \rangle$ denotes the uniform superposition state. This operation enhances the amplitude of the marked target state while suppressing non-target states, and after multiple iterations drives the measurement probability of the target state asymptotically toward unity.
\subsubsection{Measure}
In the synergistic framework of Grover's algorithm, the diffusion operator and the Oracle collaboratively amplify the probability amplitudes of target states through iterative operations.After approximately $ \frac{\pi}{4}\sqrt{\frac{N}{M}}$ optimal iterations, where $N$ denotes the size of the search space and $M$ is the number of target states, the measurement probability of the target states is boosted to nearly unity. At this stage, performing a quantum measurement causes the system to collapse into one of the target states with high probability, thereby deterministically yielding the solution to the search problem.
\subsection{The maximum clique problem(MCP)}
In an undirected graph $G(V, E)$, a clique refers to a subset $C\subseteq V $ such that for any two vertices $u$ and $v$ in $C$ , the edge $ \left ( u,v \right )\epsilon E $ . That is, the vertices in the clique are pairwise connected by edges. The maximum clique is the clique with the largest number of vertices in the graph $G$ .
For example, Fig.\ref{fig3} depicts an undirected graph defined by a vertex set $V=\left \{ v_{1}, v_{2},v_{3},v_{4},v_{5} \right \} $ and an edge set 
\begin{displaymath}
\begin{split}
E = \{ &(v_1, v_2), (v_1, v_3), (v_1, v_4), (v_1, v_5), \\
      &(v_2, v_3), (v_2, v_4), (v_3, v_4), (v_4, v_5) \}.
\end{split}
\end{displaymath}
In this graph, the cliques are 
\begin{displaymath}
\begin{split}
\mathcal{C}  = \{ \phi ,\{v_1\},\{v_2\},\{v_3\},\{v_4\},\{v_5\},\\
\{v_1, v_2\}, \{v_1, v_3\}, \{v_1, v_4\}, \{v_1, v_5\}, \\
      \{v_2, v_3\}, \{v_2, v_4\}, \{v_3, v_4\}, \{v_4, v_5\},\\
      \{v_1, v_2,v_3\},\{v_1, v_2,v_4\},\{v_1, v_3,v_4\},\\
      \{v_2, v_3,v_4\},\{v_1, v_4,v_5\},\{v_1, v_2,v_3,v_4\}
      \}.
\end{split}
\end{displaymath}
among which the largest clique in terms of vertex count is four. Therefore, the maximum clique of this graph is $\{v_1, v_2,v_3,v_4\}$.
This concept carries significant practical implications in the research of social networks.
\begin{figure}[h]
    \centering
    \includegraphics[width=0.5\textwidth]{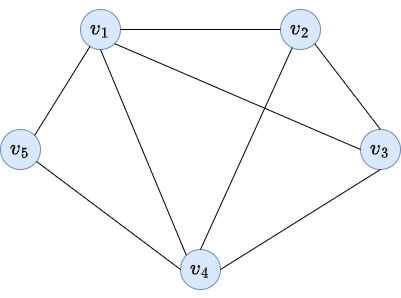}
    \caption{An undirected graph $G=\{V,E\}$ with five vertices.}
    \label{fig3}
\end{figure}
In the research context of social networks, exploring the largest group of people among whom there exists a mutual acquaintance relationship is a crucial issue. We can abstract each user in the social network as a vertex $v$ in graph theory, and the bidirectional acquaintance relationship between any two users corresponds to an undirected edge $e$. The set $V$ consisting of all vertices and the set $E$ consisting of all edges jointly define an undirected graph $G$. In this graph structure, the maximum clique of the undirected graph $G$ precisely and accurately corresponds to the largest group of people in the social network where all members know each other. The size and composition of this group directly reflect the extreme state of the closely acquainted groups within the social network.
\section{IMPLEMENTATION}\label{imple}
In this section, we propose a general framework for solving the maximum clique problem. The presented algorithm achieves identification of all maximum cliques in a graph with constant-order measurements while maintaining quantum resource consumption within reasonable bounds. Algorithm \ref{alg:max} provides the complete execution flow of our proposed approach.

\begin{algorithm}[H]
\caption{Grover's algorithm of the maximal-clique problem}\label{alg:max}
\begin{algorithmic}[1]
\Input{Adjacency matrix $adj(n,n)$ of an undirected graph $G(V,E)$}
\Output{The maximum clique of graph $G$}
\PreDetectionEncoding {Determine the possible range of values for the maximum clique size \( r \) basing on Turán's theorem and the properties of complete graphs.}
\For{$i$ in $r$}
    \State Generate all $\binom{n}{i}$ $i-vertex$ combinations
    \For{every $i$-vertex combination}
        \State Construct an $MCT$ gate controlled by the edge qubits
        \State Add the result to the counter
        \State Apply the $MCT$ gate again to reset the auxiliary qubit 
    \If{$counter = 0$}
    \State $\left | 0  \right \rangle\longrightarrow \left | info  \right \rangle_{i-1}$
\Else
    \State $\left | 1  \right \rangle\longrightarrow \left | info  \right \rangle_{i-1}$
\EndIf
    \State Use the inverse circuit to restore the counter auxiliary qubits
    \EndFor
\EndFor
\For{$j=0$ to $n-1$ , $h=0,h++$}
    \For{$p=j+1$ to $n-1$}
        \State Apply $CCCX([{\text{$ancilla1_h$}},{\text{$vertex_j$}},  {\text{$vertex_p]$},\text{$edge_{jp} $})}$
    \EndFor
\EndFor
\State Apply $MCT(ancilla1, target)$
\For{$j=0$ to $n-1$,$h=0,h++$}
    \State  Apply $Toffoli$(\text{$[target$},  \text{$vertex_j]$},\text{$ancilla2_h)$}
\EndFor
\For{$j=1$ to $n-1$} 
    \For{$h = j,h--$}
        \If{$h-1>=0$}
            \State Apply $Fredkin(\text{$vertex_j$},[\text{$ancilla2_h$}, \text{$ancilla2_{h-1}]$})$
        \EndIf
    \EndFor
\EndFor
\For{$j=0$ to $n-1$}
    \State $\left | ancilla3_j  \right \rangle  \gets \left |ancilla2_j \odot info_j \right \rangle$ 
\EndFor
\State Apply $MCT(ancilla3, out)$  
\end{algorithmic}
\end{algorithm}

Subsequently, \ref{sectionA} 
elaborates on the circuit design of each component of the algorithm, which is mainly divided into four key steps: Quantum Pre-Detection and Encoding(QPDE), Cliques Detector,  Maximal Cliques Detector and the diffusion operation, as illustrated in Fig.\ref{fig4}.
\begin{figure*}[htbp]
    \centering
    \includegraphics[width=0.9\textwidth]{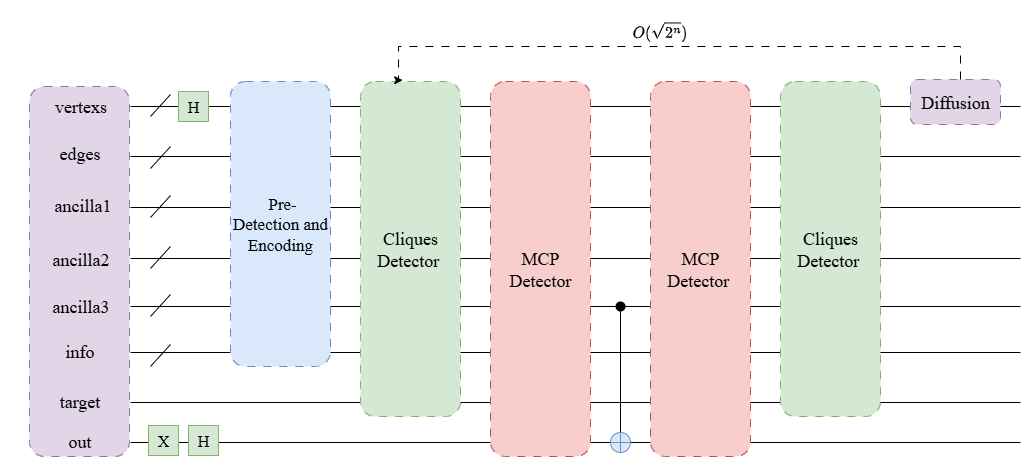}
    \caption{The framework for solving the Maximum Clique Problem. Cliques Detector, MCP Detector, and the Diffusion operation together constitute one iteration of Grover's algorithm. Here, $\left | vertexs  \right \rangle $ encodes the vertices of the graph, $\left | edges  \right \rangle  $ encodes the edges of the graph, $\left | ancilla1  \right \rangle$, $\left | ancilla2  \right \rangle $ and $\left | ancilla3  \right \rangle $ are auxiliary registers for storing intermediate values during computation, $\left | info  \right \rangle$  stores the information about the number of vertices in the maximum clique obtained during the preprocessing stage, $\left | target  \right \rangle $is used to mark the clique, and $\left | out  \right \rangle$  is used to invert the phase of the maximum clique.}
    \label{fig4}
\end{figure*}
In \ref{sectionB}, the feasibility of the algorithm is demonstrated through a case study of identifying the maximum clique in an undirected graph with four vertices.
\subsection{Algorithm Implementation Steps}\label{sectionA}
The specific implementation process of the algorithm is as follows:
\subsubsection{{Quantum Pre-Detection and Encoding}}
{To avoid the dependence of the maximum clique problem on the constant $k$, we introduce a Quantum Pre-Detection and Encoding stage.}
The main task is to obtain the vertex number information ${max_c}$ of the maximum clique in the undirected graph $G$, and pre-store the vertex number information into the quantum register $\left| x^{n} \right\rangle$. Starting from the low-order qubit $\left| x_{0} \right\rangle$ , each qubit corresponds to a positive integer. Specifically,$n$ qubits sequentially represent $n$ consecutive positive integers from 1 to $n$. When the qubit $\left| x_{i} \right\rangle$ is in the $|1\rangle$ state, it indicates that there exists a clique of size $i+1$ in the graph $G$.For instance, in the case of a graph with five vertices, the $max_c$ register requires five qubits. If, after the preprocessing stage, the quantum state of this register is transformed into $|01111\rangle$, it indicates that the maximum number of vertices in a clique within this five-vertex graph is 4. Each component state in the superposition contains the maximum clique information $max_c$, which provides a decisive condition for the subsequent labeling of maximum cliques.
\begin{enumerate}[label=(\alph*)]
    \item MCP Prior Constraints Acquisition: 
     In order to reduce the complexity of the quantum maximum clique information detection stage, we introduce the MCP Prior Constraints Acquisition stage and use Turan's theorem \cite{zhao2023graph} and the definition of the complete graph \cite{weisstein2001complete} to narrow the possible range of the maximum clique number of vertices.
     Theorem 1 and Definition 1 below elaborate on Turán's theorem and the definition of a complete graph, respectively.
\begin{theorem}[Turán's theorem]
Let $G$ be a graph with $n$ vertices. If $G$ does not contain $K_{r+1}$ (where $K_{r+1}$ denotes the complete graph on $r+1$ vertices)
as a subgraph, then the number of edges $e(G)$ satisfies:
$e(G) \leq \frac{r-1}{r} \cdot \frac{n^2}{2}$
\end{theorem}
By the contrapositive of Turán’s theorem (Theorem 1), we immediately obtain the following equivalent statement: 

\textbf{Proposition 1}. If a graph $G$ with $n$ vertices satisfies 
\begin{align}
e(G) > \frac{r-1}{r} \cdot \frac{n^2}{2}
\end{align}
then $G$ contains $K_{r+1}$ as a subgraph.

By transforming the inequality in the contrapositive statement, we obtain $r < \frac{n^2}{n^2-2e}$ . This implies that any undirected graph $G$ with $n$ vertices and $e$ edges must contain a clique of size at least\(\left\lfloor \frac{n^2}{n^2-2e} \right\rfloor + 1.\)

\begin{definition}\label{def:complete-graph}
In an undirected graph $G = (V, E)$, if there is a unique edge connecting any two distinct vertices, the graph is called a complete graph, denoted by $K_n$ , where $n$ represents the number of vertices.
\end{definition}

As shown in Definition \ref{def:complete-graph},if a clique of size $r$ exists in the graph, the number of edges $e$ must be at least $\frac{r(r-1)}{2}$, i.e., it satisfies: 
\begin{align}
  e(G) \geq \frac{r(r-1)}{2}  
\end{align}

By algebraic manipulation, the upper bound for $r$ can be derived as: $r \leq \frac{1 + \sqrt{1 + 8e}}{2}$ .Consequently, the theoretical upper bound on the size of the maximum clique in an undirected graph $G$ is given by: $\left\lfloor \frac{1 + \sqrt{1 + 8e}}{2} \right\rfloor$.This result provides an immediate combinatorial constraint for estimating the largest possible clique in the graph.

Based on the above derivation, we have clarified the structural properties of the undirected graph $G$ : there must exist a clique with vertex number $r$ satisfying $r \leq \left\lfloor \frac{n^2}{n^2 - 2e} \right\rfloor + 1$ , denoted as $l_{1} = \left\lfloor \frac{n^2}{n^2 - 2e} \right\rfloor + 1$ ; meanwhile, cliques with vertex number $r > \left\lfloor \frac{1 + \sqrt{1 + 8e}}{2} \right\rfloor$ can be strictly excluded, with $l_{2} = n - \left\lfloor \frac{1 + \sqrt{1 + 8e}}{2} \right\rfloor$.According to this rigorous theoretical boundary, we perform precise initialization on the quantum register $\left| x^{n} \right\rangle$ : its quantum state is set as $\left| x^{n} \right\rangle \to \left | 0^{l_{1} }   \right \rangle \left | x^{n-l_{1}-l_{2}  }  \right \rangle \left | 1^{l_{2} }  \right \rangle $ following the established numerical interval constraints. By setting the corresponding qubits below the lower bound and above the upper bound to 1 and 0, respectively, this operation effectively reduces the search space for subsequent quantum computations to
\begin{align}
\left\lceil \frac{n^{2}}{n^{2}-2e} \right\rceil < r \leq \left\lfloor \frac{1+\sqrt{1+8e}}{2} \right\rfloor \label{eq:r-bound}
\end{align}
 \item Quantum MCP Size Detection: In the subsequent steps, we will utilize quantum circuits to determine the existence of cliques with vertex numbers satisfying \eqref{eq:r-bound}. The results of these determinations will be stored in corresponding qubits, enabling the quantum register $\left| x^{n} \right\rangle$ to fully encode the vertex number information of the maximum clique. Specifically:
    \begin{itemize}
        \item \textbf{Vertex Set Enumeration:} 
        For each value of $r$, generate $\binom{n}{r}$ possible vertex sets, which represent all combinations of selecting $r$ vertices from $n$ vertices.
         \item \textbf{Clique Structure Determination:} 
       Using the qubits $e_{i}$ corresponding to all edges between vertices in each set as control bits, the multi-controlled Toffoli gate (MCT gate) is employed to determine whether the vertex set forms a clique. A set is classified as a clique if an edge exists between every pair of vertices within the set (i.e., all control bits are in the $\left | 1  \right \rangle $ state).
       \item \textbf{Counting and Marking:} 
      We employ a simplified quantum counter (as illustrated in Fig.\ref{fig5}
      ) to tally the number of vertex combinations that form $r$-cliques for varying values of $r$. This circuit is engineered to handle four input values, corresponding to the combinatorial number scenario where $ \binom{n}{r} = 4$.The number of counting qubits required is $\left\lfloor \log_{2} \binom{n}{r} \right\rfloor + 1$ , which corresponds to the three qubits labeled as $count_{0}$, $count_{1}$ , and $count_{2}$ in the figure. Additionally, the qubits labeled as $e^{\otimes \frac{r(r-1)}{2}}$ serve as control qubits, representing the edges between every pair of $r$ vertices. When these control qubits are all in the $\left| 1 \right\rangle^{\otimes \frac{r(r-1)}{2}}$ state, it indicates that the current vertex combination forms an $r$-clique. At this point, the input qubit of the counter labeled as $C$ will transition to the $\left| 1 \right\rangle$ state, incrementing the value of the quantum counter by 1. Subsequently, the input qubits are reset to the $\left| 0 \right\rangle$ state to facilitate the labeling of the next vertex combination.Finally, if the counter is non-zero, it indicates the presence of an $r$-clique in graph $G$. In this case, the qubit $\left| x_{r-1} \right\rangle$ is set to the $\left| 1 \right\rangle$ state.
      \begin{figure}[h]
      \centering
     \includegraphics[width=0.9\textwidth]{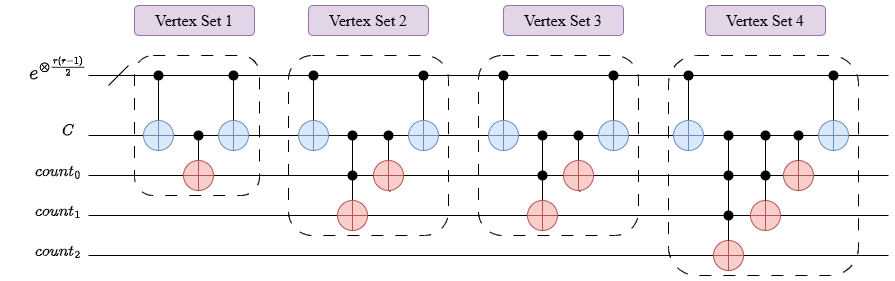}
    \caption{Quantum counter with 4 input states}
    \label{fig5}
\end{figure}
    \end{itemize}
\end{enumerate}

\subsubsection{Cliques Detector }
Before executing Cliques Detector , which is designed to identify all cliques, we need to apply Hadamard gate operations to the initial state of the quantum bits representing vertices, $\left| 0^{n} \right\rangle$ , to generate a uniform superposition state $\frac{1}{\sqrt{2^{n} } }  {\textstyle \sum_{i=0}^{2^{n}-1 }}\left | i \right \rangle  $.

In \ref{imple} above, we have elaborated on the concept of cliques. Here, we formulate the constraints for cliques in the form of conjunctive normal form (CNF). Taking a three-vertex graph as an example, the clique constraints can be expressed by the formula $(\neg A \lor \neg B \lor {AB}
) \land (\neg A \lor \neg C \lor AC) \land (\neg B \lor \neg C \lor BC)$,where the expressions connected by the conjunction symbol ($\land$) are called clauses. {Here,$A, B$, and $C$ correspond to three vertices respectively, while $AB, AC$, and $BC$ denote the edges between the corresponding pairs of vertices.}
Each clause is constructed by disjunctively combining two variables representing vertices, one variable representing their edge, and the logical operator "$\neg $".Specifically, if two vertices are simultaneously present in the set (taking the value of 1) and their incident edge does not exist in the graph (corresponding to an edge bit of 0), the clause evaluates to 0, which in turn causes the entire CNF to evaluate to 0—indicating that the current vertex set cannot form a clique. For a graph with $n$ vertices, the CNF consists of $\frac{n(n-1)}{2}$ clauses, and a vertex set can form a clique if and only if all clauses are satisfied.
For each clause, its logic is implemented using 3-controlled gates and {two X gates}
. The truth value of each clause is stored in a register labeled $\text{clause\_value}\_i$, which is initialized to the state $|1\rangle$. The $\text{clause\_value}\_i$ of the clause is flipped to $|0\rangle$ if and only if both vertex variables are $|1\rangle$ and the corresponding edge variable is $|0\rangle$ (i.e., the control bits are $|110\rangle$), as illustrated in Fig.\ref{fig6}
\begin{figure}[h]
    \centering    \includegraphics[width=0.5\textwidth]{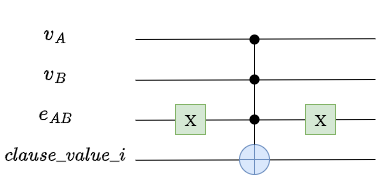}
    \caption{{Circuit to determine the truth value of the clause $\left ( \neg A\vee\neg B\vee AB  \right )$}
    .}
    \label{fig6}
\end{figure}
Finally, a multi-controlled X gate is used to implement the conjunctive operation across all clauses: this gate takes all $\text{clause\_value}$ bits as control bits. Only when all $\text{clause\_value}$ bits are $|1\rangle$ (i.e., $\left | 1^{\frac{n(n-1)}{2} }   \right \rangle $), the target bit is flipped from $|0\rangle$ to $|1\rangle$, indicating that the current vertex combination forms a clique.
\subsubsection{MCP Detector}
In Cliques Dectector
, we have successfully identified all vertex sequences in the graph that can form a clique, which are represented by the quantum states where the target qubit is in the state $|1\rangle$ .Building upon the outcomes of the preprocessing and Cliques Detector, MCP Detector is designed to identify the maximum clique and invert the phase of the corresponding quantum states, thereby preparing for the subsequent Diffusion operation. This procedure primarily encompasses three stages: Quantum state duplication, Sorting, and Consistency comparison.Fig.\ref{fig7} illustrates the designed circuit of MCP Detector for a three-vertex graph. For the Quantum state duplication and Sorting stages, we adopt the quantum algorithms developed from classical insertion sort, as presented in \cite{diasagrover}.
\begin{figure}[h]
    \centering
    \includegraphics[width=0.9\textwidth]{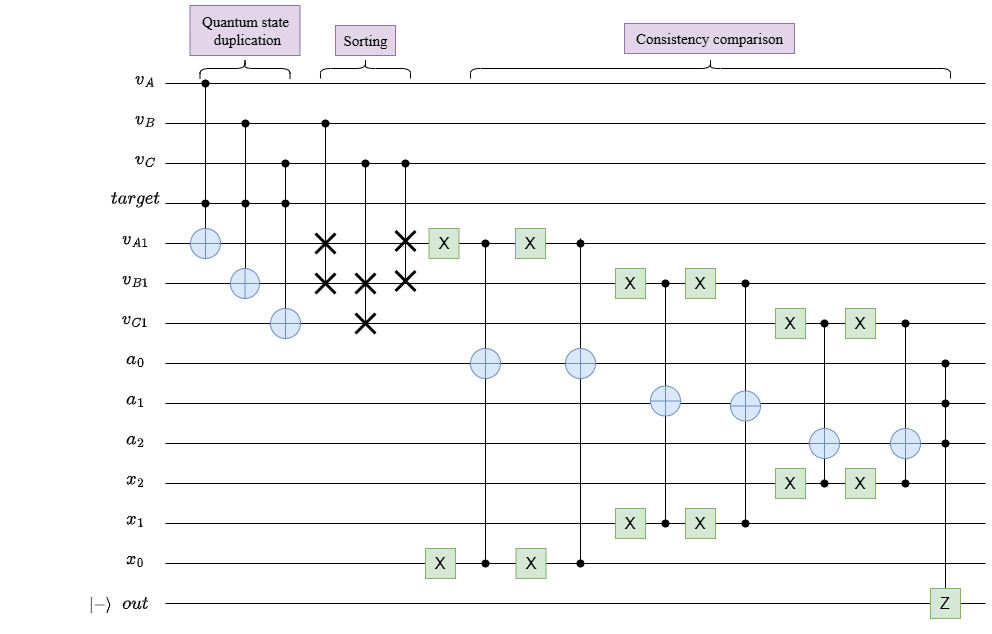}
    \caption {Quantum circuit for MCP Detector with $n = 3$ vertices to implement the phase inversion of the maximum clique.}
    \label{fig7}
\end{figure}

As shown in Fig.\ref{fig7}, the quantum state duplication stage copies the vertex quantum state information of the clique into auxiliary bits initialized to the state $|0^n\rangle$ through Toffoli gates. Sorting is performed through the Fredkin gate, whose actual effect is to move all $|1\rangle$ qubits to the end and all $|0\rangle$ qubits to the beginning. For example, the state $|101\rangle$ is transformed into $|110\rangle$. Sorting a sequence of $n$ qubits requires $\frac{n(n-1)}{2} $ Fredkin gates.In the final stage, we employ an XNOR operation to compare the sorted bit sequence with the bit sequence $\left | x_{0}x_{1}x_{2}   \right \rangle $ obtained during the preprocessing stage, which contains information about the maximum clique. If the two sequences are identical, it indicates that the current clique is indeed the maximum clique, and in this case, a Z-gate operation is applied to the output qubit labeled $\text{out}$ to invert its phase.

\subsubsection{Diffusion}
The diffusion operator achieves selective amplification of the amplitudes of target states with negative phases and the suppression of the amplitudes of non-target states by performing an "average inversion" operation on the amplitudes of all quantum states. Its mathematical expression can be described as $D = 2|s\rangle\langle s| - I$ , where $|s\rangle$ represents the uniform superposition state, i.e., $\left | s  \right \rangle =\frac{1}{\sqrt{2^{n} } }  {\textstyle \sum_{i=0 }^{2^{n}-1}} \left | i  \right \rangle$ and $n$ denotes the number of vertices in the graph for the maximum clique problem.

\subsection{Example}\label{sectionB}

To further elaborate on the above algorithm, we demonstrate a four-vertex example in Fig.\ref{fig8} ,the graph $G=(4,4)$ has $V = \{ v_{A}, v_{B}, v_{C}, v_{D} \}$ and $E = \{ e_{1} = (v_{A}, v_{B}), e_{2} = (v_{A}, v_{C}), e_{3} = (v_{A}, v_{D}), e_{4} = (v_{C}, v_{D}) \}$.The edges $e_{5} = (v_{B}, v_{D})$ and $ e_{6} = (v_{B}, v_{C})$ represented by dashed lines do not actually exist in the original graph.The clique constraints for this graph can be written as
\begin{equation}
\begin{split}
&(\neg A \lor \neg B \lor AB) \land (\neg A \lor \neg C \lor AC) \land (\neg A \lor \neg D \lor AD) \land \\
&(\neg B \lor \neg C \lor BC) \land (\neg B \lor \neg D \lor BD) \land (\neg C \lor \neg D \lor CD)
\end{split}
\label{eq:sat}
\end{equation}

\begin{figure}[h]
    \centering
    \includegraphics[width=0.4\textwidth]{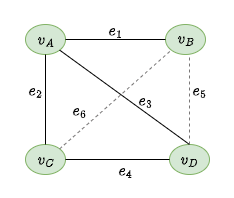}
    \caption {{Graph with four vertices.}
    }
    \label{fig8}
\end{figure}
In Fig.\ref{fig9}, we present a complete circuit diagram for solving the maximum clique problem. The qubits corresponding to $v_A$ to $v_D$ are used to encode the vertices of the graph, initialized as $\frac{1}{4} \sum_{i=0}^{15} |i\rangle$. The qubits corresponding to $e_1$ to $e_6$, which encode the edges of the graph, are initialized as $|111100\rangle$ according to Figure 8. Here,$a_0$ to $b_3$ are all auxiliary qubits, $x_3$ to $x_0$ are used to store the information about the number of vertices in the maximum clique obtained during the quantum Pre-Detection and Encoding stage, and $\text{out}$ is an output qubit for marking the maximum clique.
\begin{figure*}[htbp]
    \centering
    \includegraphics[width=\textwidth]{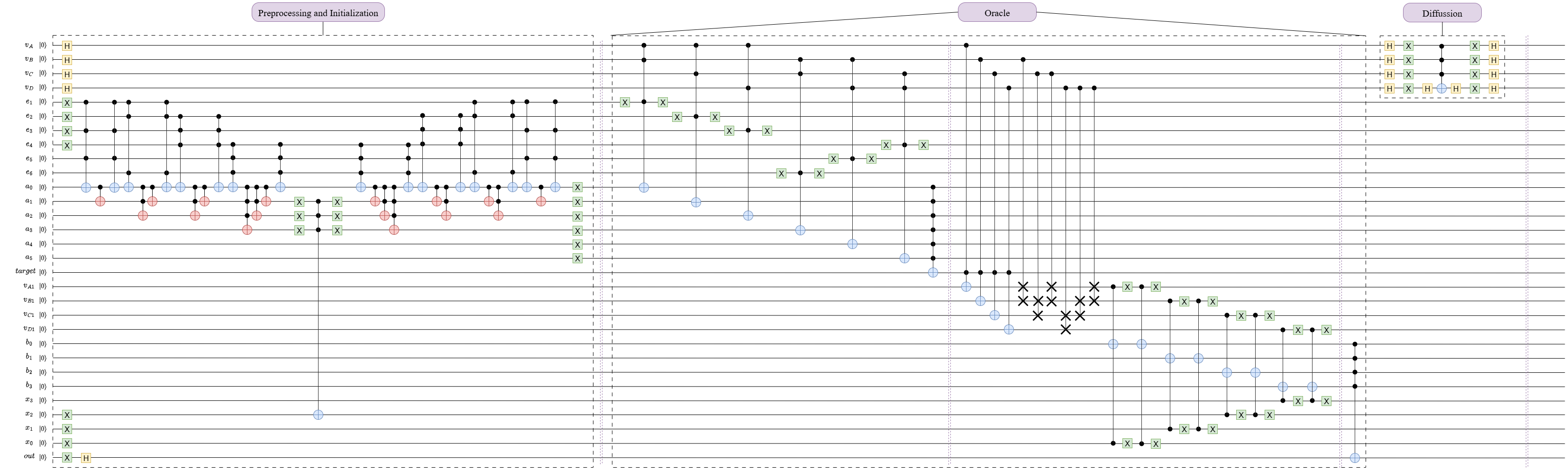} 
    \caption{This circuit is designed for solving the maximum clique problem of the four-vertex graph shown in Figure 8, and its structure is divided into two core modules: the quantum Pre-Detection and Encoding, and the Grover circuit. The Grover circuit consists of three components: Cliques Detector, MCP Detector, and the Diffusion. In practical operations, by iteratively executing the Grover circuit multiple times, the amplitude of the target state is continuously enhanced. In contrast, the amplitude of non-target states is effectively reduced, thereby improving the probability of obtaining the correct solution.}
    \label{fig9}
\end{figure*}
Before executing the quantum circuit, we obtain $2 < r \leq 3$ according to the formula \eqref{eq:r-bound} in the MCP Prior Constraints Acquisition stage of \ref{imple}. IMPLEMENTATION. It can be confirmed that the graph depicted in Figure 8 necessarily contains a clique of size 2 and does not contain a clique of size 4. The quantum state $|x_3x_2x_1x_0\rangle$ is initialized as $|0a11\rangle$ , where the $x_2$ qubit is used to characterize the existence of a complete subgraph of 3 vertex. After probing all three-vertex combinations through the quantum circuit, the state evolves to $|0111\rangle$, after which the auxiliary bits are restored. Additionally, an X-gate operation is applied to the $|a_0\ldots a_5\rangle$ register to initialize it as $|1\ldots1\rangle$, preparing for the subsequent Cliques Detector perations.

The Cliques Detector circuit is constructed based on formula \eqref{eq:sat}, which evaluates each clause and stores the results in the register $|a_0 \ldots a_5\rangle$ . If the state of this register remains $|1 \ldots 1\rangle$ , the target qubit is set to $|1\rangle$ , indicating that the current vertex combination forms a clique. Subsequently, MCP Detector copies the vertex states corresponding to the target qubit $|1\rangle$ into the register$|v_{A1} \ldots v_{D1}\rangle$ . Through the sorting operation described earlier, the $|1\rangle$ states are shifted to the most significant positions. A series of XNOR operations then compare the sorted result against the state $|0111\rangle$ , with each comparison result stored in the register $|b_0 \ldots b_3\rangle$ . When $|b_0 \ldots b_3\rangle = |1 \ldots 1\rangle$ , a negative phase is applied to the output qubit, thereby marking the quantum state corresponding to the maximum clique. The inverse of this oracle is then applied to reset the auxiliary qubits, followed by a diffusion operator to amplify the amplitudes of marked states. For the graph structure shown in Fig.\ref{fig8}, the oracle and diffusion operations are iterated four times.Finally, the result of the measurement of $|v_A v_B v_C v_D\rangle$ is shown in Fig.\ref{fig10}.
\begin{figure*}[htbp]
    \centering
    \includegraphics[width=\textwidth]{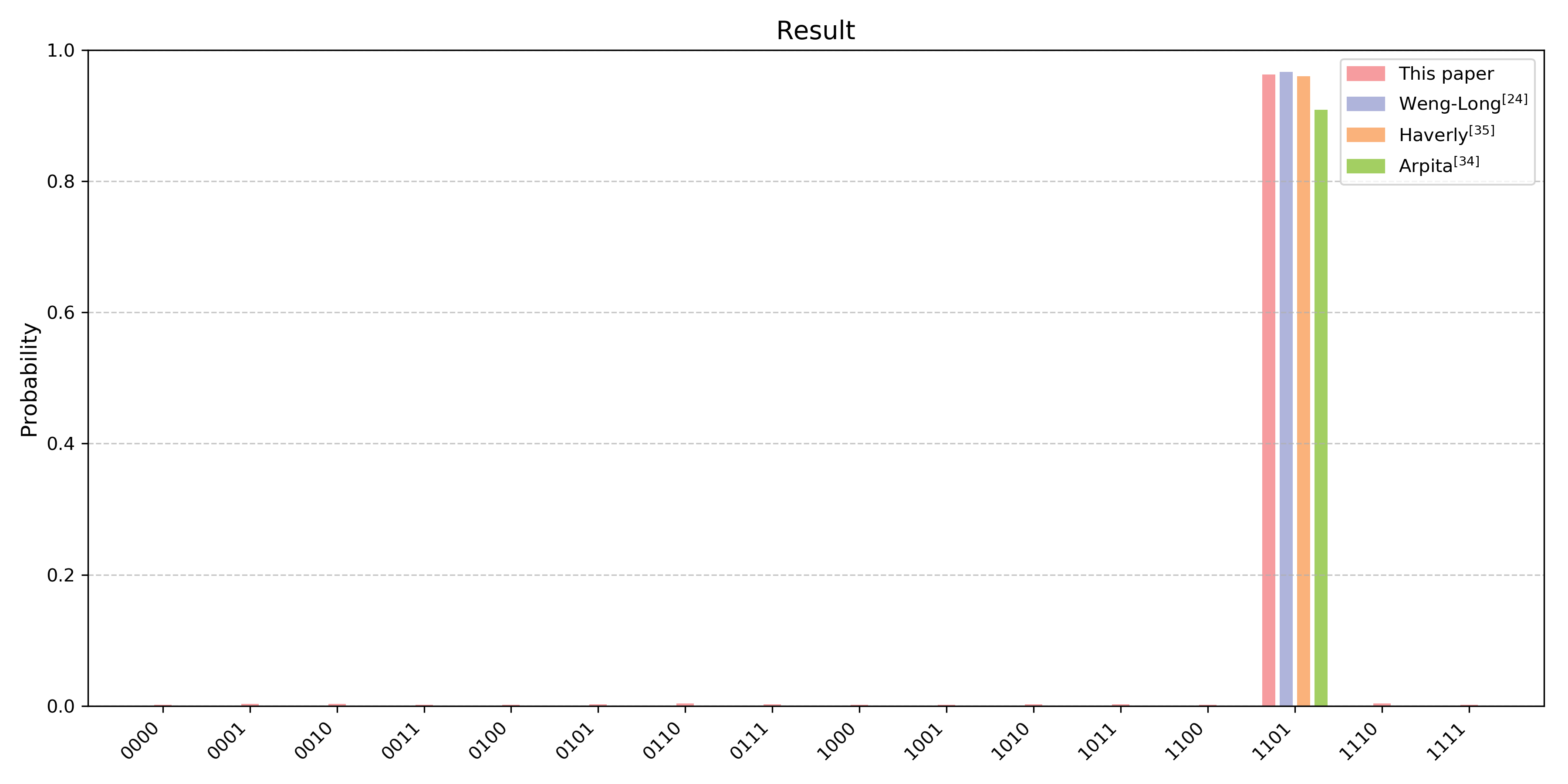}
    \caption {Simulation results for maximum clique search on the graph in Fig. 8, using the Grover-based quantum circuit shown in Fig. 9, confirm the unique solution is the vertex combination $v_Av_Cv_D$, corresponding to the binary encoding of 1011. The figure also presents the success rate distributions of different algorithms (i.e., "This paper" representing our proposed algorithm, along with the "Weng-Long", "Haverly", and "Arpita" algorithms) for the graph in Fig.\ref{fig8}. It can be observed that our method, like the other Grover-based schemes, achieves a high success rate, stabilizing near $96\%$.
    }
    \label{fig10}
\end{figure*}

\section{Results and Analysis}\label{res}
To gain deeper insights into the resource consumption of the algorithm, we analyzed the costs of the quantum circuit in terms of both the number of qubits and the number of quantum gates.
Given our introduction of a Quantum Pre-Detection and Encoding stage to address the limitations of constant $k$ dependence, excessive measurement counts, and high iteration complexity in existing methods, as well as the problem-specific redesign of the oracle, it becomes imperative to conduct a separate complexity analysis for both the preprocessing and oracle stages in addition to evaluating the overall computational overhead.
Additionally, we conducted a comparative analysis across five dimensions: the number of qubits, the number of quantum gates in Oracle operations, the total Grover iterations, the number of measurements and total quantum gates used in search for existing Grover-based approaches to the problem of maximum clique (including our method). The detailed results are presented in Table \ref{table1} .It is worth noting that we have uniformly decomposed the multiple control gates in each algorithm into a combination of single control gates and T-gates for comparison. This standardized transformation eliminates the problem of inconsistent benchmarks caused by differences in the number of control bits and establishes an equivalent reference frame for comparing the complexity of quantum circuits.
\begin{sidewaystable}
    \centering
    \caption{{Cost analysis of algorithm}}
    \label{table1}
    \renewcommand{\arraystretch}{1.5} 
    \begin{tabular}{cccccc}
        \toprule
        Method & Qubit cost & Oracle Gate cost  & Number of Grover's iterations & Measurement count & All Gates\\
        \midrule
        Haverly & $O(n^2)$ & $O(\sqrt{n}2^{n}) $ & $O(n2^{n})$ & $O(n)$ & $O(kn^{\frac{n}{2} } 2^{2n}) $ \\
        Arpita Sanyal & $O(n^2)$ & $O(\sqrt{n}2^n)$ & $O(n\sqrt{2^{n\log_{2}{n}}})$ & $O(n)$ & $O(k\sqrt{n}2^{n+\frac{n\log_{2}{n}}{2} }  )$ \\
        Matheus Giovanni Dias & $O(n^2)$ & $O(n^2)$ & $O((n\log_{2}{n})^2\sqrt{2^n})$ & $O(\log_{2}{n})$ & $O(kn^{2} (\log_{2}{n})2^{\frac{n}{2} }   )$ \\
        Weng-Long Chang & $O(n^2)$ & $O(n^2)$ & $O(n\sqrt{2^n})$ & $O(n)$ & $O(kn^{2}2^{\frac{n}{2} }   )$ \\
        This paper & $O(n^2)$ & $O(n^2)$ & $O(\sqrt{2^n})$ & $O(1)$ & $O(n^{2}2^{\frac{n}{2} } +n^{\frac{3}{2} }2^{n}) $
        \\   
        \botrule
    \end{tabular}
    
\end{sidewaystable}

\subsection{Qubit cost}
Considering a graph $G$ with $n$ vertices, the qubit resource allocation for its quantum circuit must integrate both information storage and computational auxiliary requirements, with the total number of information qubits amounting to $\frac{n^2 + 3n}{2} + 2$ . This includes $n$ qubits for encoding vertex information, $\frac{n(n-1)}{2}$ qubits for encoding edge information,$n$ qubits for storing the maximum clique vertex count obtained in the quantum Pre-Detection and Encoding stage, 1 qubit for marking clique validity, and 1 qubit for phase inversion of the maximum clique to enable amplitude amplification, thereby ensuring efficient representation of graph structures and facilitating subsequent quantum operations essential for algorithmic functionality.In the quantum Pre-Detection and Encoding stage, auxiliary qubits primarily serve the function of implementing counters, where verifying the existence of an r-clique requires $\left \lfloor \log_{2}{\binom{n}{r} }  \right \rfloor +2$ auxiliary qubits. The combinatorial number reaches its maximum when $ r=\left \lfloor \frac{n}{2}  \right \rfloor $, leading to auxiliary qubit count of $\left \lfloor \log_{2}{\binom{n}{n/2} }  \right \rfloor +2 $, which belongs to the class of complexity $ O(n)$. In the oracle stage, $\frac{n(n-1)}{2}$ qubits are required to store clause truth values,$n$ qubits to replicate clique vertex states, and $n$ qubits to hold consistency comparison results, totaling $\frac{n^2 + 3n}{2}$ auxiliary qubits and constituting an $O(n^2)$ level complexity that dominates the auxiliary resource requirements of the entire circuit. Through a qubit reuse strategy, the total number of qubits required for the entire quantum circuit is $n^2 + 3n + 2$. 

In Table \ref{table1}, we compare the qubit requirements of different Grover-based algorithms for solving the maximum clique problem, all of which exhibit $O(n^2)$ complexity in terms of quantum resource consumption. Specifically, for a graph with $n$ vertices and $m$ edges:
the Haverly algorithm requires $(n + 1)\log_{2}{n} + \frac{3n^{2}}{2} - \frac{n}{2}$ qubits;
the Arpita algorithm necessitates $k\log_{2}{k} + \frac{k^{2}}{2} - \frac{k}{2}$ qubits;
the Matheus algorithm demands $2n^{2} + 2n$ qubits; while
the Weng-Long algorithm requires $2m + n + 2 + \frac{n(n + 3)}{2}$ qubits.Notably, the qubit consumption of this work, Haverly, and Matheus is graph-density-independent. This stems from their reliance on the edge count of a complete graph (requiring $O(n^2)$ qubits) for storing edge information, rather than the actual graph's edge count ($m$). This work demonstrates an efficiency advantage over Haverly and Matheus, requiring fewer qubits. Although Arpita achieves the lowest qubit count, its vertex encoding method ($k \log_{2}{k}$) expands the solution space, increasing Grover iterations and reducing solution success probability. In contrast, the Weng-Long algorithm's qubit count is graph-density-dependent, as its edge qubits depend on the complement graph's edge count. This makes it potentially more suitable for dense graphs. However, when the complement graph's edge count approaches $O(n^2)$, its qubit requirement still exceeds that of this work.

\subsection{Number of Quantum Gates}
Since optimizing the number of quantum gates is one of the core objectives in quantum algorithm design, the number of quantum gates is adopted as the second metric for this algorithm. In the circuit, apart from single-qubit gates and single-controlled gates, multi-controlled NOT gates (MCT gates), Fredkin gates, and Toffoli gates are used more frequently. According to the property of unitary operators that any unitary operator can be implemented by single qubit gates and single controlled gates, we decompose Toffoli gates, MCT gates and Fredkin gates into combinations of single qubit gates and single controlled gates. Finally, the total number of single-qubit gates (primarily T gates obtained from Toffoli gate decomposition) and single-controlled gates is used to characterize the gate complexity.Literature \cite{nielsen2010quantum} indicates that the $C^n(U)$ gate can be implemented by using $2(n - 1)$ Toffoli gates and one controlled U gate, with the specific structure shown in Fig.\ref{fig11:subfig1}. Furthermore, the Toffoli gate can be realized through a combination of 6 T gates, 7 CNOT gates, Hadamard gates, and phase gates, the circuit structure of which is shown in Fig.\ref{fig11:subfig2}. \cite{smolin1996five}, it is shown that the Fredkin gate can be decomposed into seven single controlled gates, as illustrated in Fig.\ref{fig11:subfig3}.

\begin{figure}[htbp]
    \centering

    \subfigure[C$^n$(U) Gates]{
        \label{fig11:subfig1}
        \includegraphics[width=0.5\linewidth]{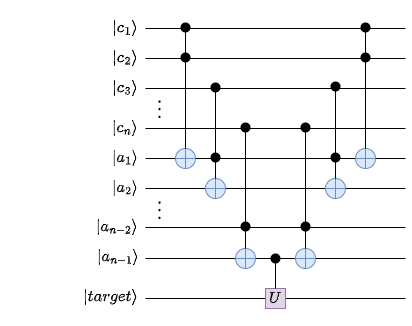}
    }
    \vspace{0.5em}
    
    \subfigure[Toffoli Gates]{%
        \label{fig11:subfig2}%
        \includegraphics[width=0.5\linewidth]{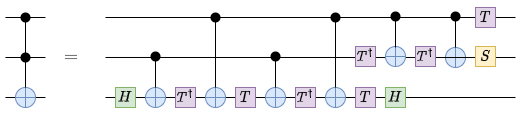}%
    }
    \vspace{0.5em}%
    
    \subfigure[Fredkin Gates]{%
        \label{fig11:subfig3}%
        \includegraphics[width=0.5\linewidth]{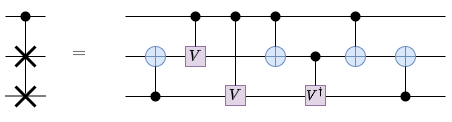}%
    }
     \caption{Quantum Gate Decompositions}
    \label{fig11}
\end{figure}
The quantity of quantum gates can be discussed in two parts: the Quantum Pre-Detection and Encoding stage and the Oracle operation. For a graph $G$ with $n$ vertices and $e$ edges, the value range of the number of vertices $r$ to be probed by the quantum circuit during the Quantum Pre-Detection and Encoding stage can be determined according to Equation (3). Therefore, a total of $\textstyle \sum_{i\in r} \binom{n}{i}$ MCT gates will be generated with the number of control bits being $\frac{i(i-1)}{2}$ to determine whether each combination of vertex forms a clique. Considering the reuse of a marker bit, we require $2 \cdot \sum_{i \in r} \dbinom{n}{i}$ MCT gates.
\begin{multline}
num_{1}= 2\sum_{i\in r} \binom{n}{i} \, (\text{MCT gate}) \\
\approx 2\sum_{i \in r} 2 \left( \frac{i(i-1)}{2} - 1 \right) \binom{n}{i} \, (\text{Toffoli gate}) \\
\approx 12 \sum_{i \in r} 2 \left( \frac{i(i-1)}{2} - 1 \right) \binom{n}{i}(\text{CNOT gate}) \\
+ 14 \sum_{i \in r} 2 \left( \frac{i(i-1)}{2} - 1 \right) \binom{n}{i}(\text{T gate})
\label{eq:5}
\end{multline}
In the counter, for each $i \in r$ , the number of 1-qubit controlled NOT gate can be derived as $\dbinom{n}{i}$, the number of 2-qubits controlled NOT gate as $\dbinom{n}{i} - 2^0$ , the number of 3-controlled NOT gate as $\dbinom{n}{i} - 2^0 - 2^1$ , and so on, until the number of $\lfloor \log_2 \dbinom{n}{i} \rfloor + 1$-controlled NOT gates is given by $\dbinom{n}{i} - 2^0 - 2^1 - \dots - 2^{\lfloor \log_2 \dbinom{n}{i} \rfloor}$. This pattern allows us to generalize the following quantitative relationships:
\begin{equation}
\begin{aligned}
\text{$C^{1}(X)$:} &\quad \sum_{i \in r} \binom{n}{i} \\
\text{$C^{2}(X)$:} &\quad \sum_{i \in r} \binom{n}{i} - (2^{2-1} - 1) \\
\text{$C^{3}(X)$:} &\quad \sum_{i \in r} \binom{n}{i} - (2^{3-1} - 1) \\
\vdots \\
\text{$C^{\lfloor \log_2 \binom{n}{i} \rfloor}(X)$:} &\quad \sum_{i \in r} \binom{n}{i} - \left(2^{\lfloor \log_2 \binom{n}{i} \rfloor} - 1\right)
\end{aligned}
\end{equation}
Based on the analysis of the equivalence relations in Fig.\ref{fig11}, the numbers of CNOT gates and T gates required for the counter can be derived respectively, and the results correspond to equation \eqref{eq:7} and equation \eqref{eq:8}:
\begin{equation}
\begin{aligned}
6 \sum_{i \in r} \Bigg[ & \binom{n}{i} - (2^{1}-1) + 2 \cdot 2 \cdot \left( \binom{n}{i} - (2^{2} - 1) \right) + \dots \\
& + 2 \cdot \left\lfloor \log_{2} \binom{n}{i} \right\rfloor \cdot \left( \binom{n}{i} - \left(2^{\left\lfloor \log_{2} \binom{n}{i} \right\rfloor} - 1\right) \right) + \\
\binom{n}{i} \Bigg]
\end{aligned}
\label{eq:7}
\end{equation}

\begin{equation}
\begin{aligned}
7 \sum_{i \in r} \Bigg[ & \binom{n}{i} - (2^{1}-1) + 2 \cdot 2 \cdot \left( \binom{n}{i} - (2^{2} - 1) \right) + \dots \\
& + 2 \cdot \left\lfloor \log_{2} \binom{n}{i} \right\rfloor \cdot \left( \binom{n}{i} - \left(2^{\left\lfloor \log_{2} \binom{n}{i} \right\rfloor} - 1\right) \right)  \Bigg]
\end{aligned}
\label{eq:8}
\end{equation}
From equations \eqref{eq:5},\eqref{eq:7} and \eqref{eq:8}, the number of gates required in the Quantum Pre-Detection and Encoding stage is derived as follows:

\begin{equation}
\begin{split}
\mathcal{N}_{\text{p}} = \sum_{i \in r} 13 *\Bigg[ & \binom{n}{i} - (2^{1}-1) + 4 \left( \binom{n}{i} - 3 \right) + \dots \\
& + 2 \left\lfloor \log_{2} \binom{n}{i} \right\rfloor \left( \binom{n}{i} - \left(2^{\left\lfloor \log_{2} \binom{n}{i} \right\rfloor} - 1\right) \right) \\
&  + 2i(i-1) \binom{n}{i} \Bigg]+ 6*\binom{n}{i}
\end{split}
\label{eq:9}
\end{equation}
In accordance with equation \eqref{eq:9}, the complexity of the number of quantum gates required in the Quantum Pre-Detection and Encoding stage is determined to be $O\left(n^{\frac{3}{2}} 2^{n}\right)$.

In the Oracle phase, we need to employ $\frac{n(n-1)}{2}$ three-controlled NOT gates and a $\frac{n(n-1)}{2}$-controlled NOT gate to complete the marking of all cliques. Meanwhile,$2n$ Toffoli gates and $\frac{n(n-1)}{2}$ Fredkin gates are utilized to achieve the screening of the maximum clique.The time complexity of the quantum gate operations in the entire oracle circuit is  $O(n^2)$.

The introduction of the QPDE phase enables direct encoding of quantum maximum clique information into the quantum circuit without prior knowledge of its specific value. This approach significantly reduces both the number of Grover iterations and measurement operations. In contrast, the other algorithms listed in Table \ref{table1} require specifying a fixed clique size constant $k$. These algorithms must iteratively adjust $k$ through repeated executions of Grover search and measurement cycles until the maximum clique is found. Crucially, in the worst case, the number of such detection operations scales as $O(n)$. Table \ref{table1} details the gate count complexity at the Oracle stage, the number of Grover iterations, and the total gate count complexity for the entire search process for each algorithm. Notably, as required detection operations 
$k$ approach 
$n$, the proposed algorithm achieves lower gate complexity than both Haverly's and Arpita's. Conversely, while Matheus and Weng-Long reduce gate complexity via 
$k
$ and edge-dependent encoding schemes, they incur higher qubit consumption.

To provide a more accurate assessment of practical resource consumption, we computed the number of qubits and the average gate count for the complete search process across all graph instances with vertex count $n \in [3, 23]$ and edge count $e \in [1, n(n-1)/2]$. Figures \ref{fig:subfig1} and \ref{fig:subfig2} compare the qubit requirements and average gate counts during the full search process between the proposed algorithm and the other four. The results demonstrate marked advantages for the proposed algorithm in quantum resource utilization efficiency.
The experimental results indicate that while the algorithms proposed by Matheus and Weng reduce the number of quantum gates, they require significantly more quantum bit resources. The algorithm by Arpita exhibits a distinct resource imbalance characteristic, achieving reduced quantum bit usage at the expense of substantially increased quantum gate count. In comparison, the Haverly algorithm shows notably higher resource consumption than our approach in both quantum gate and quantum bit dimensions. These comparative results adequately validate the performance of our algorithm in optimizing the balance of quantum computing resources.

Regarding success probability, theoretically, once the maximum clique size $k$ is determined by the Haverly, Arpita, Matheus, or Weng-Long algorithms (each $k$ adjustment cycle involves $\frac{\pi}{4} \sqrt{\frac{N}{M}}$ iterations and $O(1)$ measurements),  the success probability for finding the maximum clique, given $k$, is equivalent between these algorithms and our approach. However, the Arpita algorithm’s $\log_{2}{k}$ vertex encoding expands the solution space size $N$. This inflated $N$ directly reduces the amplitude amplification efficiency, resulting in a marginally lower success probability during its $k$-clique search. This confirms that except for the specific case in Fig. \ref{fig10}, our algorithm achieves success rates comparable to other Grover-based methods across diverse graph structures.

\begin{figure}[htbp]
    \centering
    
    \subfigure[Qubit count comparison]{
        \label{fig:subfig1}
        \includegraphics[width=0.8\linewidth]{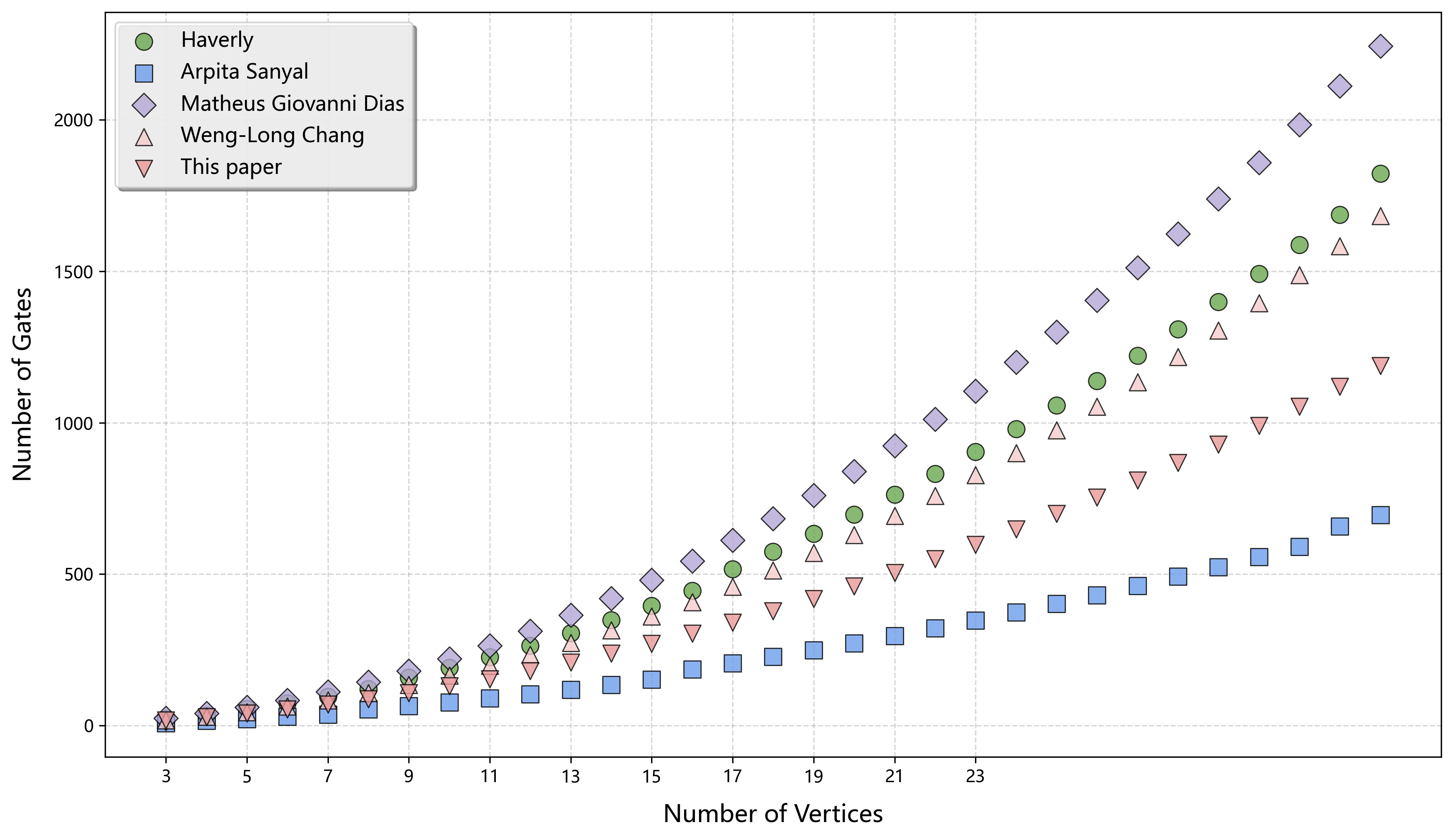}
    }
    \vspace{0.5em}
    
    \subfigure[Average gate count comparison]{
        \label{fig:subfig2}
        \includegraphics[width=0.8\linewidth]{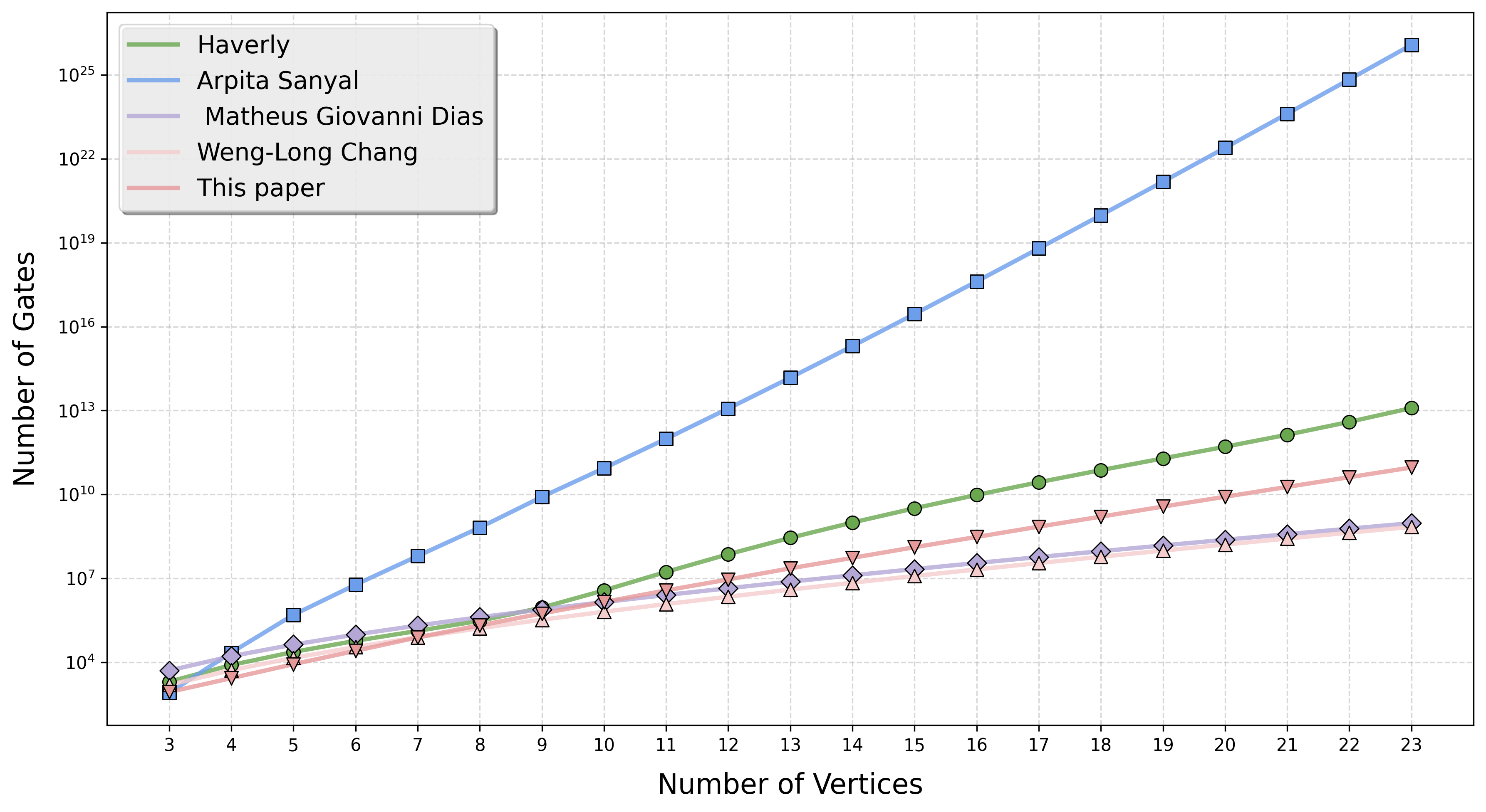}
    }
     \caption{Comparison of quantum resource consumption between the proposed algorithm and four existing Grover-based maximum clique algorithms during the full search process.}
    \label{fig:mainfig}
\end{figure}
\section{Conclusion}\label{con}
This paper presents a general solution to the maximum clique problem utilizing Grover's algorithm. By incorporating the concept of global variables, we introduce a Quantum Pre-Detection and Encoding stage that synergizes with Circuits Detector and MCP Detector, enabling efficient identification of the maximum clique with $O(1)$ measurements and $O(\sqrt{2^n})$ Grover iterations, while maintaining a controlled number of qubits and quantum gates.

Furthermore, we conduct a comparative analysis with existing Grover-based maximum clique algorithms, specifically evaluating their respective requirements for qubit count, quantum gate complexity, and success probability.Our study demonstrates that while the Quantum Pre-Detection and Encoding (QPDE) stage exhibits a gate complexity of $O(n^\frac{3}{2}2^{n})$, this procedure is executed only once during the entire search process. Consequently, its contribution to the overall quantum resource overhead is mitigated. The total gate complexity is $O(n^{2}2^{\frac{n}{2} } +n^{\frac{3}{2} }2^{n})$, which remains competitive among comparable methods.
And the algorithm proposed in this study  achieves a success rate comparable to the current state-of-the-art Grover-based maximum clique algorithms. While maintaining this advantage,our algorithm also demonstrates the
following improvements in quantum resource consumption:
the worst-case quantum bit complexity is $O(n^2 + 3n + 2)$,
which is superior to other existing schemes.  To validate the proposed quantum circuit, we performed simulations on the 4-vertex graph model shown in Fig. \ref{fig8} using Python and IBM's Qiskit framework. Results confirm that the circuit outputs the maximum clique vertex set with high probability, achieving a $96\%$ success rate comparable to other Grover-based methods, thereby verifying its effectiveness and reliability. Nevertheless, despite its single execution mitigating prohibitive overall gate count growth, the QPDE stage still contributes significantly to the total complexity. Future work will therefore explore alternative QPDE strategies to reduce this computational burden. Complementing this effort, optimizing qubit consumption remains a priority; we will investigate advanced encoding strategies and circuit design optimization techniques to further minimize quantum resource overhead.

\bmhead{Acknowledgements}

 This research was supported by the National Natural Science Foundation of China (Grant No. 62101600) and the State Key Lab of Processors, Institute of Computing Technology, CAS (Grant No. CLQ202404).

\bibliography{sn-bibliography}
\end{document}